\documentstyle[aps,multicol,epsf]{revtex}
\input epsf
\draft
\title{Matter wave interference using two-level atoms and resonant optical fields}
\author{B. Dubetsky and P. R. Berman 
\\ {\em Physics Department, University of Michigan, Ann Arbor,\\
Michigan 48109-1120} }
\date{\today }
\begin{document}
\maketitle
\begin{abstract}
A theory of matter wave interference is developed in which resonant,
standing wave optical fields interact with an ensemble of two-level atoms.
If effects related to the recoil the atoms undergo on absorbing or emitting
radiation are neglected, the {\em total} atomic density is spatially
uniform. However, when recoil effects are included, spatial modulation of
the atomic density can occur for times that are greater than or comparable
with the inverse recoil frequency, $\omega _{q}^{-1}$. In this regime, the
atoms exhibit matter-wave interference that can be used as the basis of a
matter wave atom interferometer. Two specific atom field geometries are
considered. In the first, atoms characterized by a homogeneous velocity
distribution are subjected to a single radiation pulse. The pulse excites
the atoms which then decay back to the lower state. The spatial modulation
of the total atomic density is calculated as a function of $\ t$, where $t$
is the time following the pulse. In contrast to the normal Talbot effect,
the spatially modulated density is {\em not }a periodic function of $\ t,$
owing to spontaneous emission; however, after a sufficiently long time, the
contribution from spontaneous processes no longer plays a role and the
Talbot periodicity is restored. In the second atom-field geometry, there are
two pulses separated by an interval $T$. The atomic velocity distribution in
this case is assumed to be inhomogeneously broadened. Owing to the
inhomogeneous broadening, one finds a nonvanishing spatial modulation of the
density only at specific ''echo times'' following the second pulse. In
contrast to the normal Talbot-Lau effect, the spatially modulated density is 
{\em not }a periodic function of $T,$ owing to spontaneous emission;
however, for sufficiently long time, the contribution from spontaneous
processes no longer plays a role and the Talbot periodicity is restored. The
structure of the spatially modulated density in the vicinity of the echo
times is studied, and is found to mirror the atomic density following the
first pulse. With a suitable choice of observation time and field strengths,
the spatially modulated atomic density serves as an indirect probe of the
distribution of spontaneously emitted radiation.
\end{abstract}


\section{Introduction}

Atom interferometry has emerged as an important new research area in the
past fifteen years \cite{rev}. Thermal atoms offer unique properties as the
working element of an interferometer, owing to their small de Broglie
wavelength. Atom interferometry has already led to impressive results in
precision measurements of $\hbar $ \cite{Chu}, atomic polarizability \cite
{polar}, rotation rates \cite{9,10} and the acceleration of gravity \cite
{Chu1}. Applications to nanolithography have also been proposed \cite{42}.
Despite the advances in atom interferometry, there remains some controversy
as to what constitutes an atom interferometer. We have discussed previously 
\cite{8} the distinction between classical and quantum scattering for atom
interferometers in which microfabricated gratings are used as the
''optical'' elements in the interferometer. In this paper, we extend this
discussion to atom interferometers that employ optical fields to modify the
atomic state wave function.

In our effort to classify atom interferometers, it will prove useful to
summarize the conclusions we reached in analyzing the scattering of atoms by
microfabricated gratings \cite{8}. As a model system, we consider the
scattering of an atomic beam by two microfabricated gratings separated by a
distance $L$. The internal structure of the atoms is irrelevant here, since
the atoms are in their ground state and the scattering by the gratings is
assumed to be the same for any and all ground state sublevels - there is no
internal state coherence. For an angular divergence of the atomic beam, $%
\theta _{b},$ satisfying $\theta _{b}\gg d/L$, where $d$ is the grating
period, the atomic density is uniform in the region between the gratings,
but can exhibit ''fringes'' beyond the second grating. In particular, at a
distance $2L$ from the first grating, fringes having period $d$ are
produced. As such, this device has the appearance of an interferometer.
Moreover, it can be used to measure inertial effects \cite{19}, since the
fringes are displaced if the atoms accelerate between the interaction zones.
Further evidence that this device constitutes an interferometer is provided
by a theoretical description of the scattering in which the microfabricated
gratings create coherence between transverse momentum components of the
incoming atomic beam, taken as a plane wave. Despite all these features, it
is incorrect to conclude that such a system legitimately qualifies as an
atom interferometer.

To clarify matters, it is best to back up a bit and consider the scattering
of an atomic beam by a single grating. At a distance $L$ from the grating,
the nature of the scattering is determined by the so called Talbot length 
\cite{Talbot}, defined as 
\[
L_{T}=2d^{2}/\lambda _{dB}, 
\]
where $\lambda _{dB}=h/mu_{0}$ is the atomic de Broglie wavelength ($m$ is
the atomic mass and $u_{0}$ is the longitudinal atomic speed). For $L\ll
L_{T}$, diffraction effects are unimportant and one is in the classical
scattering limit. For a highly collimated beam ($\theta _{b}\ll d/L$), the
pattern produced on a screen at a distance $L\ll L_{T}$ from the
microfabricated grating is simply the {\em shadow} of the grating. This is
clearly a classical effect.

Of course, the classical scattering limit can be obtained from a calculation
in which the center-of-mass motion is quantized. If such a fully quantum
calculation is carried out, one finds that, immediately following the
grating, the atomic state wave function contains coherence between different
transverse momentum components. The {\em atomic state amplitude grating} or
matter grating created by the microfabricated grating necessarily implies
the existence of such coherence. However, for $L\ll L_{T}$, this coherence
is {\em not} evidence for quantum scattering since it corresponds simply to
the classical shadow effect. The relative phase associated with the
coherence associated with two transverse momentum components, $p$ and $%
p+\hbar q$, with $\hbar q\equiv 2\pi \hbar /d,$ can be written as 
\[
\phi =q\frac{p}{m}t+\omega _{q}t, 
\]
where $t=L/u_{0}$ and $\omega _{q}=\hbar q^{2}/2m$ is a recoil frequency.
The first term is a Doppler shift, independent $\hbar $ - it has a classical
origin. The second term is a recoil shift which vanishes in the limit $\hbar
\sim 0$. It is a purely quantum term. Thus, for $\omega _{q}t\ll 1$, which
is equivalent to the condition $L\ll L_{T},$ the scattering is classical and
the coherence between transverse momentum components corresponds to a
classical Doppler shift. It should now be clear that the fringes produced in
the two-grating interferometer have nothing to do with quantization of the
atomic motion when $L\ll L_{T}$. They result simply from a shadow effect and
constitute a moir\'{e} pattern. For grating separations $L\gtrsim L_{T}$,
quantum diffraction effects begin to play a role; however, the conditions
for fringe production are still determined by classical considerations
related to cancellation of a Doppler phase.

We return now to atom interferometers which use optical fields to create
atomic coherence between internal states, external states, or both.
Depending on which coherence is created and probed, one can classify such
atom interferometers into three general categories. Classical atom
interferometers (CAI) in no way rely on quantization of the atoms'
center-of-mass motion for their operation. The CAI operate in the regimes $%
\omega _{q}t\ll 1$ or $L\ll L_{T}$, for which quantum scattering effects can
be ignored. If one restricts the use of the term ''atom interferometer'' to
those devices which depend on quantization of the atoms' center-of-mass
motion for their operation, the CAI would not qualify. Examples of CAI
include those based on radio-frequency \cite{Ramsey} or optical \cite
{Bergquist,Vasilenko,18} Ramsey fringes.

Matter wave atom interferometers operate only in the regime $\omega
_{q}t\gtrsim 1$ or $L\gtrsim L_{T}.$ We distinguish matter wave atom
interferometers in which the {\em total} atomic density is monitored from
those in which an internal state coherence or population is probed. The
former we refer to simply as matter wave atom interferometers (MWAI) and the
latter as internal state matter wave atom interferometers (ISMWAI). An
important feature which distinguishes MWAI from ISMWAI is the dependence of
the observed signal on the parameter $\omega _{q}t$ or $L/L_{T}$. The ISMWAI
signal does not vanish in the limit $\omega _{q}t\ll 1$ or $L\ll L_{T}$, but 
{\em does} vanish in this limit for MWAI. In other words, MWAI depend
critically on the quantization of the atoms' center of mass motion for their
operation, whereas ISMWAI can operate in the classical or quantum scattering
limits. Although matter wave effects can modify the signals in ISMWAI, many
features of the signals are determined by classical considerations, as in
CAI. Examples of ISMWAI interferometers include those which explore recoil
splitting of optical Ramsey fringes in the frequency domain \cite
{Dubetsky,Barger,Helmcke,Borde,Chu,Ertmer}, and time-domain \cite{Chu1,16,17}%
.

Matter wave atom interferometers can be constructed using off-resonant
standing wave optical fields. Such fields act as phase gratings for the
matter waves. Atoms enter a field interaction region in their ground state
and leave the region with their spatial density unaffected. On the other
hand, the {\em phase} of the ground state wave function is spatially
modulated, corresponding to a coherence between momentum states differing by 
$n\hbar q$ ($n$ is an integer and $4\pi /q$ is the period of the standing
wave field). As the matter wave evolves freely following the field
interaction region, the phase modulated wave function is transformed into a
spatially modulated density, which can be probed using various techniques.
Atom interferometers can be constructed using two or more field interaction
zones \cite{12p,12,Bragg}. In the transient echo experiment of Cahn {\it et
al. } \cite{12}, the vanishing of spatial modulation at the echo times $t=2T$
and $t=3T$ (but not in the immediate vicinity of the echo times) is
consistent with MWAI theory \cite{2}. Cahn {\it et al. }also found that the
spatial modulation vanished in the classical scattering region, $\omega
_{q}T\ll 1$, as it must for any MWAI.

It is also possible to create MWAI by resonantly driving one-photon
transitions between electronic states \cite{2} or two-photon Raman
transitions between ground state sublevels using optical fields. The
lifetime of the electronic or ground state coherence must be greater than $%
\omega _{q}^{-1},$ a condition that restricts one to intercombination lines
in the case of electronic state coherence. It is possible to show \cite{2}
that the sum of the spatially modulated atomic gratings created by the
fields in each of the states involved in the transition (that is, the total
atomic density) vanishes in the classical scattering limit, but is
nonvanishing in the quantum scattering domain, $\omega _{q}t\gtrsim 1$. This
effect has not yet been observed. If one were to probe the total atomic
density rather than some internal state density in the experiments of \cite
{Barger,Helmcke,Ertmer}, the sought after effect should be observable.

In this paper, we describe a new type of MWAI \cite{Yakovlev} that also uses
resonant rather than off-resonant fields. Two-level atoms pass through a
standing wave, optical field that resonantly drives a {\em closed} two-level
transition in the atoms. In contrast to the scheme described above, it is
now assumed that the excited state lifetime is much less than $\omega
_{q}^{-1}$. As a result of the atom-field interaction, an excited state
matter grating in the excited state is created, as is a ''hole grating'' in
the ground state. If one neglects all effects related to quantization of the
atoms' center-of-mass motion, it is easy to show that, after the excited
state decay back to the ground state, the ground state density is uniform.
In some sense, the excited state grating fills the ''hole'' in the ground
state density that had been created by the field. When the recoil associated
with stimulated \cite{20} and spontaneous\cite{21,Baklanov} processes are
included, however, new features appear in the ground state density matrix.
It is not possible to interpret the interaction in simple terms as was the
case for a phase grating, but the net result is that recoil leads to a
nonvanishing spatially modulated atomic density for $\omega _{q}t\gtrsim 1$
or $L\gtrsim L_{T}$. In analyzing the signal, it will prove useful to
separate the contributions from spontaneous and stimulated processes.
Immediately following the excited state decay, these two contributions
cancel one another, but as time progresses, the contribution from
spontaneous processes disappears, leaving a net modulated ground state
density. When two field interaction zones are used, echo-like phenomena can
occur. It will be seen that {\em both} stimulated and spontaneous processes
contribute to the echo signals, even for pulse separations $T\gg \omega
_{q}^{-1}.$

The paper is arranged as follows. The change in the atomic density matrix
following the interaction of atoms with a standing wave, optical field is
calculated in Sec. II. In Sec. III we consider atom interference using a
single atom-field interaction zone and a homogeneous distribution of atomic
velocities. It is seen that focussing of the atoms, similar to that found
with phase gratings, can also occur using resonant fields. The conditions
under which the spatially modulated atomic density becomes a periodic
function of $t$ (Talbot effect) are established. The atomic density
following two atom-field interaction zones is calculated for an
inhomogeneously broadened transition (Talbot-Lau regime) in Sec. IV. The
results are summarized in Sec. V.

\section{Basic equations}

Atom interferometers can operate in the spatial or time domain. In the
spatial domain an atomic beam traverses one or more field regions. In the
time domain, a vapor of cold atoms (or condensate) is subjected to one or
more radiation pulses. The spatial domain interferometer can be analyzed in
the time domain if calculations are carried out in the atomic rest frame.
Consequently, we restrict our calculations to the time domain.

Two-level (upper state $\left| e\right\rangle $ and ground state $\left|
g\right\rangle $) atoms are subjected to two radiation pulses separated by a
time interval $T$. {\em Each} radiation pulse consists of two traveling wave
components having propagation vectors ${\bf k}_{1}$ and ${\bf k}_{2}$,
respectively, $\left| {\bf k}_{1}\right| =\left| {\bf k}_{2}\right|
=k=\Omega /c,$ and $\Omega $ is the frequency of each field. The total
electric field can be written as 
\begin{equation}
E\left( {\bf r},t\right) ={\bf \hat{e}}e^{-i\Omega t+i{\bf Q\cdot r}}\cos
\left( {\bf q\cdot r/}2\right) \left[ E_{1}g_{1}(t)+E_{2}g_{2}(t-T)\right]
+c.c.,  \label{1}
\end{equation}
where ${\bf Q=}\left( {\bf k}_{1}+{\bf k}_{2}\right) /2,\,\,{\bf q=}\left( 
{\bf k}_{1}-{\bf k}_{2}\right) ,$ ${\bf \hat{e}}$ is a polarization vector, $%
E_{j}$ is the amplitude of pulse $j$ $\left( j=1,2\right) $, and $%
g_{j}\left( t\right) $ is a smooth pulse envelope function having width $%
\tau $, centered at $t=T_{j},$ with $T_{1}=0$ and $T_{2}=T.$ We assume that
the pulse separation $T,$ pulse duration $\tau ,$ atom-field detuning $%
\Delta =\Omega -\omega $ ($\omega $ is the atomic transition frequency),
upper state decay rate $\Gamma ,$ recoil frequency 
\begin{equation}
\omega _{q}=\hbar q^{2}/2m,  \label{1p}
\end{equation}
and Doppler shift ${\bf k}_{i}\cdot {\bf v}$ (${\bf v}$ is an atomic
velocity) satisfy the inequalities 
\begin{mathletters}
\label{2}
\begin{eqnarray}
\Gamma T &\gg &1,  \label{2a} \\
\Gamma \tau &\ll &1,  \label{2b} \\
\Delta \tau &\ll &1,  \label{2c} \\
{\bf k}_{i}\cdot {\bf v} &\ll &\Gamma ,  \label{2d} \\
\omega _{q}T &\geq &1.  \label{2e}
\end{eqnarray}
Inequality (\ref{2a}) implies that any excited state population created by
the fields decays to the ground state in a time that is short compared with
the time scale of the experiment. Inequalities (\ref{2b}, \ref{2c}) imply
that spontaneous decay and atom-field detuning can be neglected during the
radiation pulses, while inequality (\ref{2d}) (atoms cooled below the
Doppler limit of laser cooling) guarantees that there is negligible Doppler
dephasing for times of order of the excited state lifetime. Finally
condition (\ref{2e}) states that we are in the quantum scattering limit.

Conditions (\ref{2}) allow one to map out the time development of the
density matrix resulting from each radiation pulse in three stages:

\begin{enumerate}
\item  an impulsive change in the density matrix produced by the atom-field
interaction

\item  spontaneous decay of the excited state

\item  free evolution of the density matrix.
\end{enumerate}

\noindent Depending on the specific application, we calculate the atomic
density matrix following the first or second pulse. Owing to inequalities (%
\ref{2}), we can define times before ($T_{j}^{\left( -\right) })$ and
following ($T_{j}^{\left( +\right) })$ pulse $j$ such that changes in the
atomic density resulting from spontaneous decay and atom-field detunings can
be neglected in the time interval $T_{j}^{\left( +\right) }-T_{j}^{(-)}$.

\subsection{Stage 1}

During pulse $j$, the density matrix, in an interaction representation,
evolves according to 
\end{mathletters}
\begin{equation}
i\hbar \dot{\rho}=\left[ V,\rho \right] ,  \label{3}
\end{equation}
\begin{equation}
V=2\hbar \chi _{j}g_{j}\left( t-T_{j}\right) \cos \left( {\bf q\cdot r/}%
2\right) \left[ \cos \left( {\bf Q\cdot r}\right) \sigma _{x}-\sin \left( 
{\bf Q\cdot r}\right) \sigma _{y}\right] ,  \label{4}
\end{equation}
where $\chi _{j}=-\mu E_{j}/2\hbar $ is a Rabi frequency, $\mu $ is a dipole
moment operator matrix element, $\sigma _{x}$ and $\sigma _{y}$ are Pauli
matrices, and 
\begin{equation}
\left| e\right\rangle =\left( 
\begin{array}{c}
1 \\ 
0
\end{array}
\right) ,\,\,\left| g\right\rangle =\left( 
\begin{array}{c}
0 \\ 
1
\end{array}
\right) .  \label{5}
\end{equation}
It has been assumed that, during each pulse, any effects arising from atomic
motion can be neglected [inequalities (\ref{2b}, \ref{2d})], which is the
reason the kinetic energy term has been omitted from the Hamiltonian in Eq. (%
\ref{3}) (Raman-Nath approximation). Before the first pulse acts, it is
assumed that all atoms are in their ground state. Owing to inequality (\ref
{2a}), all population is returned to the ground state before the action of
the second pulse at $t=T_{2}$. Thus we need only calculate the change in the
ground and excited state density matrices produced by pulse $j$, starting
from a density matrix 
\begin{equation}
\rho \left( {\bf r,r}^{\prime };T_{j}^{\left( -\right) }\right) =\left( 
\begin{array}{cc}
0 & 0 \\ 
\rho _{gg}\left( {\bf r,r}^{\prime };T_{j}^{\left( -\right) }\right) & 0
\end{array}
\right) ,  \label{6}
\end{equation}
in which all atoms are in their ground state. One can integrate Eq. (\ref{3}%
) to obtain the density matrix immediately following the pulse at time $%
T_{j}^{\left( +\right) },$%
\begin{mathletters}
\label{7}
\begin{eqnarray}
\rho \left( {\bf r,r}^{\prime };T_{j}^{\left( +\right) }\right) &=&\eta
_{j}\left( {\bf r}\right) \rho \left( {\bf r,r}^{\prime };T_{j}^{\left(
-\right) }\right) \eta _{j}^{\dagger }\left( {\bf r}^{\prime }\right) ,
\label{7a} \\
\eta _{j}\left( {\bf r}\right) &=&\cos \left[ \,^{1}/_{2}\theta _{j}\cos
\left( {\bf q\cdot r/}2\right) \right] -i\sin \left[ \,^{1}/_{2}\theta
_{j}\cos \left( {\bf q\cdot r/}2\right) \right] \left[ \cos \left( {\bf %
Q\cdot r}\right) \sigma _{x}-\sin \left( {\bf Q\cdot r}\right) \sigma _{y}%
\right] ,  \label{7b}
\end{eqnarray}
where 
\end{mathletters}
\begin{equation}
\theta _{j}=4\chi _{j}\int_{-\infty }^{\infty }dtg_{j}\left( t\right)
\label{7p}
\end{equation}
is a pulse area.

It is convenient to use the Wigner representation for the density matrix, 
\[
\rho \left( {\bf r,p},t\right) =\int \frac{d{\bf \hat{r}}}{\left( 2\pi \hbar
\right) ^{3}}\exp \left( -i{\bf p\cdot \hat{r}/}\hbar \right) \rho \left( 
{\bf r+\hat{r}/2,r-\hat{r}/2},t\right) , 
\]
and expand the populations as

\begin{equation}
\rho _{nn}\left( {\bf r,p},t\right) =\sum_{s}\rho _{nn}\left( s,{\bf p}%
,t\right) \exp \left[ is{\bf q\cdot r}\right] .  \label{9}
\end{equation}
Using Eqs. (\ref{7})-(\ref{9}) and expanding the $\sin \left[
\,^{1}/_{2}\theta _{j}\cos \left( {\bf q\cdot r/}2\right) \right] $ and $%
\cos \left[ \,^{1}/_{2}\theta _{j}\cos \left( {\bf q\cdot r/}2\right) \right]
$ functions in terms of Bessel functions, one obtains for the Fourier
coefficients 
\begin{mathletters}
\label{10}
\begin{eqnarray}
\rho _{gg}\left( s,{\bf p},T_{j}^{\left( +\right) }\right) &=&\sum_{\ell
,s^{\prime }}\left( -1\right) ^{s^{\prime }}J_{2\ell }\left( \theta
_{j}/2\right) J_{2\left( \ell -s^{\prime }\right) }\left( \theta
_{j}/2\right) \rho _{gg}\left[ s-s^{\prime },{\bf p-}\hbar {\bf q}\left(
\ell -s^{\prime }/2\right) ,T_{j}^{\left( -\right) }\right] ,  \label{10a} \\
\rho _{ee}\left( s,{\bf p},T_{j}^{\left( +\right) }\right) &=&\sum_{\ell
,s^{\prime }}\left( -1\right) ^{s^{\prime }}J_{2\ell +1}\left( \theta
_{j}/2\right) J_{2\left( \ell -s^{\prime }\right) +1}\left( \theta
_{j}/2\right) \rho _{gg}\left\{ s-s^{\prime },{\bf p-}\hbar \left[ {\bf q}%
\left( \ell -\left( s^{\prime }-1\right) /2\right) +{\bf Q}\right]
,T_{j}^{\left( -\right) }\right\} ,  \label{10b}
\end{eqnarray}
where $J_{s}(x)$ is a Bessel function of order $s$.

\subsection{Stages 2 and 3}

In the next stages of the calculation, the excited state decays to the
ground state and the ground state density matrix evolves freely following
the decay. The Wigner function associated with the excited state population
obeys the equation of motion 
\end{mathletters}
\begin{equation}
\left( \frac{\partial }{\partial t}+{\bf v\cdot \nabla }\right) \rho
_{ee}\left( {\bf r,p},t\right) =-\Gamma \rho _{ee}\left( {\bf r,p},t\right)
\label{11}
\end{equation}
It then follows from Eq. (\ref{11}) that the $s$-order Fourier component
evolves as 
\begin{equation}
\rho _{ee}\left( s,{\bf p},t\right) =\exp \left[ -\left( \Gamma +is{\bf %
q\cdot p/}m\right) \left( t-T_{j}\right) \right] \rho _{ee}\left( s,{\bf p}%
,T_{j}^{\left( +\right) }\right)  \label{12}
\end{equation}
The excited state repopulates the ground state. The ground state Wigner
function is governed by the equation 
\begin{equation}
\left( \frac{\partial }{\partial t}+{\bf v\cdot \nabla }\right) \rho
_{gg}\left( {\bf r,p},t\right) =\Gamma \int d{\bf n}_{r}N\left( {\bf n}%
_{r}\right) \rho _{ee}\left( {\bf r},{\bf p+}\hbar {\bf k}_{r},t\right) ,
\label{121}
\end{equation}
from which it follows that the ground state Fourier coefficients evolve as 
\begin{equation}
\left( \frac{\partial }{\partial t}+is{\bf q\cdot p/}m\right) \rho
_{gg}\left( s{\bf ,p},t\right) =\Gamma \int d{\bf n}_{r}N\left( {\bf n}%
_{r}\right) \rho _{ee}\left( s,{\bf p+}\hbar {\bf k}_{r},t\right)
\label{122}
\end{equation}
In these equations, ${\bf k}_{r}$ is a spontaneous photon wave vector, ${\bf %
n}_{r}={\bf k}_{r}/k_{r},$ and $N\left( {\bf n}_{r}\right) $ is the
normalized probability density for the radiation of a photon in the
direction ${\bf n}_{r}$.

The solution of Eq. (\ref{121}) involves both a homogeneous and particular
solution which we write as 
\begin{equation}
\rho _{gg}\left( s{\bf ,p},t\right) =\rho _{gg}^{\left( S\right) }\left( s,%
{\bf p},t\right) +\rho _{gg}^{\left( D\right) }\left( s,{\bf p},t\right) .
\label{14}
\end{equation}
The homogeneous part, 
\begin{equation}
\rho _{gg}^{(S)}\left( s,{\bf p},t\right) =\exp \left[ -is{\bf q\cdot p}%
\left( t-T_{j}\right) /m\right] \rho _{gg}\left( s,{\bf p},T_{j}^{\left(
+\right) }\right) ,  \label{10p}
\end{equation}
represents the evolution of the ground state in the absence of decay. The $S$
superscript indicates that this part is associated purely with stimulated
processes. The particular solution, $\rho _{gg}^{\left( D\right) }\left( s,%
{\bf p},t\right) ,$ represents the contribution to the ground state density
matrix resulting from excitation by the pulse and subsequent decay of the
excited state (hence, the superscript $D$ for decay). One finds that at
times $t-T_{j}\gg \Gamma ^{-1}$ the particular solution is given by 
\begin{equation}
\rho _{gg}^{\left( D\right) }\left( s,{\bf p},t\right) =\exp \left[ -is{\bf %
q\cdot p}\left( t-T_{j}\right) /m\right] \int d{\bf n}_{r}N\left( {\bf n}%
_{r}\right) \left( 1+is\omega _{d}n_{q}/\Gamma \right) ^{-1}\rho _{ee}\left(
s,{\bf p+}\hbar {\bf k}_{r},T_{j}^{\left( +\right) }\right) ,  \label{13}
\end{equation}
where 
\begin{equation}
\omega _{d}=\hbar qk_{r}/m  \label{13p}
\end{equation}
and $n_{q}={\bf n}_{r}{\bf \cdot q/}q.$ The quantity $s\omega _{d}$ is a
recoil frequency associated with the spontaneous decay of the $s$th excited
state Fourier component. Alternatively, $s\omega _{d}$ can be viewed as a
Doppler shift of the spontaneously emitted photon that is dependent on the
momentum kick $s\hbar q$ which the atom acquires in the excitation process.

Piecing together Eqs. (\ref{13}), (\ref{10p}), (\ref{10}), one finds that
the ground state density matrix for $t-T_{j}\gg \Gamma ^{-1}$ is given by 
\begin{eqnarray}
\rho _{gg}\left( s,{\bf p},t\right) &=&\sum_{\ell ,s^{\prime }}\left(
-1\right) ^{s^{\prime }}\exp \left[ -is{\bf q\cdot p}\left( t-T_{j}\right) /m%
\right] \left\{ J_{2\ell }\left( \theta _{j}/2\right) J_{2\left( \ell
-s^{\prime }\right) }\left( \theta _{j}/2\right) \rho _{gg}\left[
s-s^{\prime },{\bf p-}\hbar {\bf q}\left( \ell -s^{\prime }/2\right)
,T_{j}^{\left( -\right) }\right] \right.  \nonumber \\
&&+J_{2\ell +1}\left( \theta _{j}/2\right) J_{2\left( \ell -s^{\prime
}\right) +1}\left( \theta _{j}/2\right) \int d{\bf n}_{r}N\left( {\bf n}%
_{r}\right) \left( 1+is\omega _{d}n_{q}/\Gamma \right) ^{-1}  \nonumber \\
&&\times \left. \rho _{gg}[s-s^{\prime },{\bf p-}\hbar \left( {\bf q}\left(
\ell -s^{\prime }/2\right) +{\bf Q-k}_{r}\right) ,T_{j}^{\left( -\right)
}]\right\}  \label{13pp}
\end{eqnarray}
This is the building-block solution which can be used to analyze several
possible experimental schemes.

For $\omega _{d}^{-1},\omega _{q}^{-1}\gg t-T_{j}\gg \Gamma ^{-1}$, recoil
effects can be neglected and the terms involving recoil momenta in the
arguments of the density matrix elements can be dropped. In that limit the
sum over $\ell $ of the Bessel functions gives $\delta _{s^{\prime },0}$,
and Eq. (\ref{13pp}) reduces to $\rho _{gg}\left( s,{\bf p},t\right) =\exp %
\left[ -is{\bf q\cdot p}\left( t-T_{j}\right) /m\right] \rho _{gg}(s,{\bf p}%
,T_{j}^{\left( -\right) })$, which implies that 
\begin{equation}
\rho _{gg}\left( {\bf r},{\bf p}\text{,}t\right) =\rho _{gg}\left( {\bf r-}%
\frac{{\bf p}}{m}(t-T_{j}),{\bf p,}T_{j}^{\left( -\right) }\right) .
\label{131}
\end{equation}
As expected, the density matrix simply undergoes a classical translation if
recoil is neglected.

\section{One Field-Interaction Zone}

When an electromagnetic wave passes through an amplitude or phase grating,
the diffraction pattern as a function of the distance from the grating is a
periodic function of the Talbot length, or, equivalently, of $\omega _{q}t$.
Similar effects occur for matter waves and have been observed experimentally
by Chapman et al \cite{23} and Nowak et al \cite{Mlyneck}. In these
experiments, one sent ground state atoms through a microfabricated grating.
When atoms are excited by resonant, optical fields, the diffraction pattern
is no longer strictly periodic owing to spontaneous decay. In this section
we calculate the atomic density following the interaction of a highly
collimated atomic beam ($qu\theta _{b}t\ll 1)$ or a condensate ($qut\ll 1$)
with a resonant optical field. In the atomic rest frame, the field appears
as a pulse centered at $T_{1}=0.$ The initial Wigner distribution is taken
as $\rho _{gg}\left( {\bf r},{\bf p},T_{1}^{\left( -\right) }\right)
=W\left( {\bf p}\right) $ \cite{normal}$,$ corresponding to Fourier
components 
\begin{equation}
\rho _{gg}\left( s,{\bf p},T_{1}^{\left( -\right) }\right) =\delta
_{s,0}W\left( {\bf p}\right) .  \label{132}
\end{equation}
In this equation, $W\left( {\bf p}\right) $ is the momentum distribution of
the atoms.

It is convenient to orient the $x,\,\,y,\,\,z\,\,$\ axes\ along the mutually
orthogonal vectors ${\bf q,\,\,Q,\,\,}$and ${\bf k}_{1}\times {\bf k}_{2.}$
It then follows from Eqs. (\ref{13pp}), along with the summation identities 
\begin{mathletters}
\label{19}
\begin{eqnarray}
S_{\nu }^{\left( e\right) }\left( a\right) &\equiv &\sum_{\ell }J_{2\ell
}\left( a\right) J_{2\left( \ell -\nu \right) }\left( a\right) \exp \left(
-i\ell \alpha \right)  \nonumber \\
&=&\frac{1}{2}\exp ^{-i\nu \alpha /2}\left\{ J_{2\nu }\left[ 2a\cos \left(
\alpha /4\right) \right] +\left( -1\right) ^{\nu }J_{2\nu }\left[ 2a\sin
\left( \alpha /4\right) \right] \right\} ,  \label{19a} \\
S_{\nu }^{\left( o\right) }\left( a\right) &\equiv &\sum_{\ell }J_{2\ell
+1}\left( a\right) J_{2\left( \ell -\nu \right) +1}\left( a\right) \exp
\left( -i\ell \alpha \right)  \nonumber \\
&=&-\,\frac{1}{2}\exp ^{-i\left( \nu -1\right) \alpha /2}\left\{ J_{2\nu }%
\left[ 2a\cos \left( \alpha /4\right) \right] -\left( -1\right) ^{\nu
}J_{2\nu }\left[ 2a\sin \left( \alpha /4\right) \right] \right\} ,
\label{19b}
\end{eqnarray}
that the Fourier component at time $t,$ 
\end{mathletters}
\begin{equation}
\rho _{gg}\left( s,t\right) =\int d{\bf p}\rho _{gg}\left( s,{\bf p,}%
t\right) ,  \label{16}
\end{equation}
is given by 
\begin{eqnarray}
\rho _{gg}\left( s,t\right) &=&\,^{1}/_{2}\left\langle \exp \left( -is{\bf %
q\cdot p}t/m\right) \right\rangle \left\{ J_{2s}\left[ \theta _{1}\sin
\left( \phi _{T_{S}}\left( st\right) /2\right) \right] \right. \left[
1+C\left( \phi _{T_{D}}\left( st\right) ,s\omega _{q}/\Gamma \right) \right]
\nonumber \\
&&+(-1)^{s}J_{2s}\left[ \theta _{1}\cos \left( \phi _{T_{S}}\left( st\right)
/2\right) \right] \left. \left[ 1-C\left( \phi _{T_{D}}\left( st\right)
,s\omega _{d}/\Gamma \right) \right] \right\} ,  \label{17}
\end{eqnarray}
where $\left\langle \ldots \right\rangle $ represents an average over atomic
momenta, 
\begin{equation}
\phi _{T_{S}}\left( t\right) =\omega _{q}t  \label{18a}
\end{equation}
is the Talbot phase associated with stimulated processes, 
\begin{equation}
\phi _{T_{D}}\left( t\right) =\omega _{d}t  \label{18e}
\end{equation}
is the Talbot phase associated with spontaneous processes, and 
\begin{equation}
C\left( \alpha ,\beta \right) =\int d{\bf n}N\left( {\bf n}\right) \exp
\left( i\alpha n_{x}\right) \left( 1+i\beta n_{x}\right) ^{-1}.  \label{18d}
\end{equation}
Terms involving $C\left( \alpha ,\beta \right) $ in Eq. (\ref{17}) are
connected with spontaneous processes while the remaining terms arise from
stimulated processes.

It follows from Eq. (\ref{17}) that the spatially homogeneous part of the
atomic density is unchanged, 
\begin{equation}
\rho _{gg}\left( 0,t\right) =1,  \label{180}
\end{equation}
consistent with probability conservation in the closed two-level system. We
are interested primarily in the time dependence of the Fourier components
having $s\neq 0$, since these components determine the spatial modulation of
the atomic density. The maximum value of $s$ entering the summation in Eq. (%
\ref{9}) is of order $\max \left\{ 1,\theta _{1}\right\} $. It is assumed in
this section that the Doppler broadening is small, 
\begin{equation}
\max \left\{ 1,\theta _{1}\right\} {\bf q\cdot p}t/m\ll 1,  \label{181}
\end{equation}
which allows one to set $\left\langle \exp \left( -is{\bf q\cdot p}%
t/m\right) \right\rangle $ equal to unity in Eq. (\ref{17}). Note that, even
without this factor, the general expression (\ref{17}) is {\em not }a
periodic function of time. Owing to spontaneous emission the Talbot effect
is destroyed. We will see below that, for sufficiently large times, the
Talbot periodicity is restored.

If both traveling wave components of the field are linearly-polarized along%
{\bf \ }${\bf z}$ and if the ground state angular momentum is equal to $0,$
then 
\begin{equation}
N\left( {\bf n}\right) =\,^{3}/_{8\pi }\left( 1-n_{z}^{2}\right) .
\label{20}
\end{equation}
We assume that the recoil frequency is sufficiently small to ensure that 
\begin{equation}
\max \left\{ 1,\theta _{1}\right\} \omega _{q}/\Gamma \ll 1.  \label{21}
\end{equation}
As a consequence one need only evaluate (\ref{18d}) at $\beta =0$. Using the
identity \cite{22p} 
\begin{equation}
\int d{\bf n}n_{i}n_{k}\exp [i{\bf a\cdot n]}=4\pi a^{-3}\left\{ \left( \sin
a-a\,\cos a\right) \delta _{i,k}+\frac{a_{i}a_{k}}{a^{2}}\left[ \left(
a^{2}-3\right) \sin a+3a\cos a\right] \right\}  \label{211}
\end{equation}
one finds 
\begin{equation}
C\left( \alpha ,0\right) =\,^{3}/_{2}\alpha ^{-3}\left[ \alpha \cos \alpha
+\left( \alpha ^{2}-1\right) \sin \alpha \right] .  \label{22}
\end{equation}
Note that $C\left( \alpha ,0\right) \sim 1-3\alpha ^{2}/16$ for $\alpha \ll
1 $ and $C\left( \alpha ,0\right) \sim \frac{3}{2\alpha }\sin \alpha $ for $%
\alpha \gg 1$.

The time scale of the transient response (\ref{17}) is determined by the
recoil frequencies, 
\begin{mathletters}
\label{22p}
\begin{equation}
\omega _{d}=4\omega _{k}\sin \left( \Theta /2\right) ,\,\,\,\,\,\,\omega
_{q}=4\omega _{k}\sin ^{2}\left( \Theta /2\right)  \label{22pb}
\end{equation}
where $\Theta $ is the angle between the wave vectors ${\bf k}_{1}$ and $%
{\bf k}_{2}.$ There are essentially two time scales in the problem, one
associated with spontaneous processes, $\tau _{d}=\omega _{d}^{-1}$, and one
associated with stimulated processes $\tau _{q}=\omega _{q}^{-1}$. For
Fourier components having $s\neq 0$, it is possible to isolate the
contribution from stimulated processes since the contribution from
spontaneous processes becomes negligible for $t\gg \tau _{d}$. It is not
difficult to understand why spontaneous processes contribute a negligible
amount in this limit. The recoil phase factor associated with spontaneous
processes \cite{Berman} is $e^{i{\bf k}_{r}\cdot {\bf q}t/2m}=e^{in_{r_{x}}%
\omega _{d}t}$. As mentioned above, this phase factor can be viewed as a
recoil related Doppler phase. When summed over all directions of the
spontaneously emitted photon, it averages to zero for $\omega _{d}t\gg 1$.
As a consequence, one finds from Eq. (\ref{17}) that, for $\omega _{d}t\gg 1$
and $s\neq 0,$ 
\end{mathletters}
\begin{equation}
\rho _{gg}\left( s,t\right) =\,^{1}/_{2}\left\{ J_{2s}\left[ \theta _{1}\sin
\left( \phi _{T_{S}}\left( st\right) /2\right) \right] +(-1)^{s}J_{2s}\left[
\theta _{1}\cos \left( \phi _{T_{S}}\left( st\right) /2\right) \right]
\right\} .\,\,  \label{222}
\end{equation}

In effect, Eq. (\ref{222}) represents a periodic rephasing (Talbot effect)
of {\em the ground state amplitude grating} that was created {\em %
immediately following} the radiation pulse, since for $\omega _{d}t\gg 1$
the spontaneous contribution no longer plays a role. One could have equally
well ionized all the excited state atoms immediately following the pulse.
The Fourier components, $\rho _{gg}\left( s,t\right) ,$ and total density, $%
\rho _{gg}\left( {\bf r},t\right) =\sum_{s}\rho _{gg}\left( s,t\right) e^{is%
{\bf q\cdot r}}$, are periodic functions having period equal to $2\pi
/\omega _{q}$ and can be used to measure recoil frequency \cite{12,ground},
but, in contrast to scattering by phase gratings, there is no time for which
the density is uniform when $\omega _{d}t\gg 1.$ When all Fourier components
are taken into account, and for large pulse areas (but not so large as to
violate the Raman-Nath approximation), 
\begin{equation}
\left( \omega _{q}\tau \right) ^{-1}\gg \theta _{1}\gg 1,  \label{223}
\end{equation}
it can be shown that the atoms are focused by the field. This new regime of
atom focusing, as well as its relation to focusing by phase gratings, will
be considered in a future publication.

\begin{figure}[tb!]
\centering
\begin{minipage}{8.0cm}
\epsfxsize= 8 cm \epsfysize= 7.7 cm \epsfbox{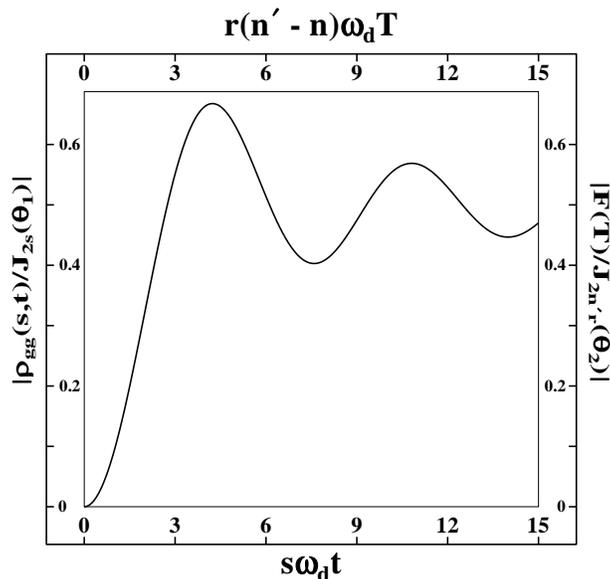}
\end{minipage}
\caption{The quantity $1-C\left( x,0\right) $ is plotted as a function of $x$%
. When $\Theta \ll 1,$ this function gives the time-dependence of the
Fourier component $\left| \protect\rho _{gg}\left( s,t\right) /J_{2s}\left( 
\protect\theta _{1}\right) \right| $ $\left( x=s\protect\omega _{d}t\right) $
for $t\sim \protect\omega _{d}^{-1}\ll \protect\omega _{q}^{-1}$ and $%
\left| F\left( T\right) /J_{2n^{\prime }r}\left( \protect\theta _{2}\right)
\right| $ $\left[ x=r\left( n^{\prime }-n\right) \protect\omega _{d}T\right] 
$ for $T\sim \protect\omega _{d}^{-1}\ll \protect\omega _{q}^{-1}.$}
\label{Fig1}
\end{figure}

For earlier times, when $t\lesssim \omega _{d}^{-1}\lesssim \omega
_{q}^{-1}, $ the spontaneous term contributes to the atomic density. If $%
t\ll \omega _{d}^{-1},$ the total density is approximately uniform since
spontaneous decay ''refills'' the ''hole'' in the ground state that is
created by the radiation pulse. From Eqs. (\ref{17}, \ref{22}), one finds
that 
\begin{equation}
\rho _{gg}\left( s,t\right) \approx \frac{1}{64}\left( s\omega _{q}t\right)
^{2}\theta _{1}^{2}\delta _{s,1}+\frac{1}{10}\left( -1\right) ^{s}\left(
s\omega _{d}t\right) ^{2}J_{2s}\left( \theta _{1}\right) .  \label{221}
\end{equation}
The Fourier components build up as $t^{2}$ when $t\ll \omega _{d}^{-1}$.
When the angle between wave vectors is small $\left( \Theta \ll 1\right) ,$
and for somewhat larger times, $t\sim \omega _{d}^{-1}\ll \omega _{q}^{-1}$,
Eqs. (\ref{17}) reduces to 
\begin{equation}
\rho _{gg}\left( s,t\right) =\,^{1}/_{2}(-1)^{s}J_{2s}\left[ \theta _{1}%
\right] \left[ 1-C\left( \phi _{T_{D}}\left( st\right) ,0\right) \right] ,
\label{224}
\end{equation}
allowing one to isolate the contribution from spontaneous processes. The
Fourier component $\rho _{gg}\left( 1,t\right) $ is plotted in Fig. \ref
{Fig1} as a function of $\omega _{d}t$; it is not a periodic function of $%
\omega _{d}t$. The time dependence of Eq. (\ref{224}) can serve as a probe
of the spontaneous emission distribution function [see Eq. (\ref{18d})].

\begin{figure}[tb!]
\centering
\begin{minipage}{8.0cm}
\epsfxsize= 8 cm \epsfysize= 7.7 cm \epsfbox{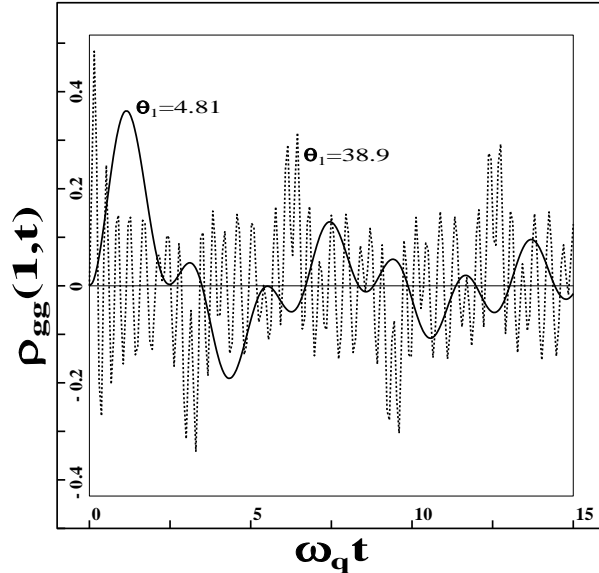}
\end{minipage}
\caption{Fourier component $\protect\rho _{gg}(1,t)$ for a single field
interaction zone. The pulse area $\protect\theta _{1}=38.9$ (dashed curve)
is chosen to maximize $\protect\rho _{gg}(1,t),$ while the area $\protect%
\theta _{1}=4.81$ (solid curve) is chosen to maximize the relative
contribution from spontaneous processes.}
\label{Fig2}
\end{figure}

For counterpropagating waves ($\Theta =\pi $), the recoil frequencies $%
\omega _{q}$ and $\omega _{d}$ coincide and achieve their maximum value $%
\omega _{d}=\omega _{q}=4\omega _{k}$. In this limit, atom interference
effects occur on the shortest possible time scale. The Fourier component $%
\rho _{gg}\left( 1,t\right) $ is plotted as a function of $\omega _{q}t$ in
Fig. \ref{Fig2} for $\Theta =\pi $. Other Fourier components could be shown
as well, but $\rho _{gg}\left( 1,t\right) $ is the component most easily
monitored using backscattering techniques. One sees that, initially, $\rho
_{gg}\left( 1,t\right) $ is aperiodic, but it asymptotically approaches
periodic behavior for $\omega _{d}t\gg 1$. The area $\theta _{1}=38.9$ is
that for which $\left| \rho _{gg}\left( 1,t\right) \right| $ achieves its
maximum. The area $\theta _{1}=4.81$ is chosen to maximize the relative
contribution of spontaneous processes. It was obtained by maximizing the
ratio $\rho _{m}/\rho _{m}^{as}$, where $\rho _{m}$ is the maximum of the
exact expression (\ref{17}), which occurs at $\omega _{q}t\sim 1$, and $%
\rho _{m}^{as}$ is the maximum of the asymptotic expression (\ref{222})
occurring at $\omega _{q}t\gg 1$.

\section{Two Field Interaction Zones}

In the previous section, it was assumed that any Doppler dephasing was
negligible on the time scales under consideration. In this section, we look
at the limit in which 
\begin{equation}
qu\gg \omega _{d},\omega _{q},  \label{225}
\end{equation}
where $u$ is the width of the velocity distribution in the direction of $%
{\bf q}$. To be specific, we take the momentum distribution as $W({\bf p}%
)=W_{\bot }({\bf p}_{\bot })W_{q}(p_{q})$, 
\begin{equation}
W_{q}(p_{q})=\frac{1}{\sqrt{\pi }mu}e^{-p_{q}^{2}/(mu)^{2}}  \label{226}
\end{equation}
is the distribution of momenta in the direction of ${\bf q}$ and $W_{\bot }(%
{\bf P}_{\bot })$ is the distribution of momenta transverse to ${\bf q}$. By
combining Eqs. (\ref{17}) and (\ref{226}) one finds that, following a single
pulse, the Fourier components of the density are given by 
\begin{equation}
\rho _{gg}\left( s,t\right) =e^{-(squt)^{2}/4}J_{2s}\left[ \theta _{1}\phi
_{T_{S}}\left( st\right) /2\right] ,  \label{227}
\end{equation}
correct to order $\omega _{d}/qu.$ This term survives only for times of
order $\left( squ\right) ^{-1}$, which implies that the argument of the
Bessel function is of order $\theta _{1}\omega _{q}/qu$. This, in turn,
implies that only Fourier components having $s\sim \max $ $\left\{
1,\theta _{1}\omega _{q}/qu\right\} $ contribute significantly for times in
which Fourier components other than $s=0$ are nonvanishing.

The picture is rather simple. The radiation pulse excites the atoms which
then decay back to the ground state giving a uniform density. For times $t<$ 
$\left( squ\right) ^{-1}$, the $s$th Fourier component begins to build up
significantly provided $\theta _{1}\omega _{q}/qu>s$, and a modulated atomic
density appears. For times $t>$ $\left( qu\right) ^{-1}$, all spatial
modulation has been washed out as a result of Doppler dephasing. Since this
time is shorter than the inverse recoil frequencies, the main features of
the time dependence $\rho _{gg}\left( s,t\right) $ found in Sec. III for
homogeneous broadening never can be realized in this inhomogeneously
broadened sample.

The spatial modulation is not lost, however, and can be restored using echo
techniques if a second pulse is applied at some time $T$ following the
initial pulse. The time dependence of $\rho _{gg}\left( s,t,T\right) $,
considered as a function of $T,$ displays the same features found in the
previous section for $\rho _{gg}\left( s,t\right) $ as a function of $t$.
Specifically, it can be used as a probe of spontaneous processes.

In this section, we consider the atomic response following two pulses
centered at $t=0$ and $t=T$ $\left( T_{1}=0\text{ and }T_{2}=T\right) $. One
key point to remember is that the relevant time window for which the
modulation is restored is of order $\left( squ\right) ^{-1}$. Thus, it is
possible that the Doppler dephasing associated with spontaneous emission no
longer plays a critical role in killing off the Fourier components, since
this dephasing is negligible on a time scale $\left( squ\right) ^{-1}$. We
shall see this to be the case; as a consequence the Fourier components at
the ''echo times'' have contributions from {\em both} the stimulated and
spontaneous terms, even for $\omega _{d}t\gg 1$.

The Fourier components of the density at times $t-T_{2}\gg \Gamma ^{-1}$
following the second pulse can be written as 
\begin{equation}
\rho _{gg}\left( s{\bf ,p,}t\right) =\rho _{gg}^{(SS)}\left( s{\bf ,p,}%
t\right) +\rho _{gg}^{\left( DS\right) }\left( s{\bf ,p,}t\right) +\rho
_{gg}^{\left( SD\right) }\left( s{\bf ,p,}t\right) +\rho _{gg}^{\left(
DD\right) }\left( s{\bf ,p,}t\right) ,  \label{23}
\end{equation}
where $\rho _{gg}^{\left( IK\right) }(s,{\bf p},t)$ represents the
contribution from stimulated ($K=S$) or spontaneous ($K=D$) processes
following the second pulse that depend on the stimulated ($I=S$) or
spontaneous ($I=D$) component of the ground state density matrix Fourier
components that were created by the first pulse$.$ Consider, for example $%
\rho _{gg}^{(SS)}\left( s{\bf ,p,}t\right) .$ Using Eqs. (\ref{10a}, \ref
{10p}, \ref{132}, \ref{226}), one finds that the Fourier component 
\begin{equation}
\rho _{gg}^{(SS)}\left( s,t\right) =\int d{\bf p}\rho _{gg}^{(SS)}\left( s,%
{\bf p,}t\right) ,  \label{24}
\end{equation}
is given by 
\begin{eqnarray}
\rho _{gg}^{(SS)}\left( s,t\right) &=&\sum_{s^{\prime }}\exp \left\{
-n^{2}q^{2}u^{2}\left[ s(t-T)+(s-s^{\prime })T\right] ^{2}/4\right\} 
\nonumber \\
&&\times \sum_{\ell _{1}}\exp \left\{ -2i\omega _{q}\left[ \ell _{1}-\left(
s-s^{\prime }\right) /2\right] \left[ s(t-T)+(s-s^{\prime })T\right]
\right\} J_{2\ell _{1}}\left( \theta _{1}/2\right) J_{2\left( \ell
_{1}-s+s^{\prime }\right) }\left( \theta _{1}/2\right)  \nonumber \\
&&\times \sum_{\ell _{2}}\exp \left[ -2is\omega _{q}\left( t-T\right) \left(
\ell _{2}-s^{\prime }/2\right) \right] J_{2\ell _{2}}\left( \theta
_{2}/2\right) J_{2\left( \ell _{1}-s^{\prime }\right) }\left( \theta
_{2}/2\right)  \label{25}
\end{eqnarray}

When the time separation between pulses is larger than \ the inverse Doppler
width, 
\begin{equation}
quT\gg 1,  \label{26}
\end{equation}
the average over momenta leads to a non-vanishing contribution only if 
\begin{equation}
s(t-T)+(s-s^{\prime })T\lesssim (qu)^{-1}.  \label{261}
\end{equation}
Inequality (\ref{261}) can be satisfied in the vicinity of the echo times, $%
t_{e},$ defined as 
\begin{equation}
t_{e}=\frac{n^{\prime }}{n}T,  \label{27}
\end{equation}
where $n^{\prime }$ and $n$ ($n^{\prime }>n)$ are positive integers having
no common factors, provided that 
\begin{equation}
s=nr\text{, }s^{\prime }=n^{\prime }r,  \label{28}
\end{equation}
where $r$ is an integer. Setting 
\[
\delta t=t-t_{e}=t-\frac{n^{\prime }}{n}T, 
\]
and using inequality (\ref{261}) and Eqs. (\ref{28}), one finds that the
Doppler phase is nondestructive for times 
\[
\delta t\lesssim 1/\left( nrqu\right) 
\]
Although not indicated explicitly, $\delta t$ is a function of $%
t,n,n^{\prime }$ and $T$.

Since $nr=s$, Eq. (\ref{28}) implies that the $(nr)$th Fourier component ($%
r=1,2,3\ldots $) is nonvanishing in the vicinity of the echo time. For
example, if $n=1$, all the Fourier components contribute near the echo times 
$t=n^{\prime }T$ ($n^{\prime }=2,3\ldots )$, corresponding to a macroscopic
atomic grating having period $\lambda /\left[ 2\sin \left( \Theta /2\right) %
\right] $; if $n=2$, the $(2r)$th Fourier components contribute near the
echo times $t=n^{\prime }T/2$ ($n^{\prime }=3,5,7\ldots )$, corresponding to
a macroscopic atomic grating having period $\lambda /\left[ 4\sin \left(
\Theta /2\right) \right] $. In this manner one can generate macroscopic
atomic gratings having period $\lambda /\left[ 2n\sin \left( \Theta
/2\right) \right] $. Note that condition (\ref{261}) for a nondestructive
Doppler phase is a {\em classical} condition since it does not contain $%
\hbar $. The shape of the echo signal about the echo times and the
dependence of the signal on the time separation of the pulses is determined
by effects related to quantum scattering, but the actual {\em location }of
the signals is determined by classical considerations only \cite{8}.

Using Eq. (\ref{19a}) for the sums over $\ell _{1}$ and $\ell _{2},$ one
finds that, at $\delta t=t-(n^{\prime }/n)T$, 
\begin{eqnarray}
\rho _{gg}^{(SS)}\left( nr,\delta t\right)  &=&\,^{1}/_{4}\exp \left[
-(nrqu\delta t)^{2}/4\right]   \nonumber \\
&&\times \left\{ J_{2r\left( n^{\prime }-n\right) }\left[ \theta _{1}\sin
\left( \phi _{T_{S}}\left( nr\delta t\right) /2\right) \right] +\left(
-1\right) ^{\left( n^{\prime }-n\right) r}J_{2r\left( n^{\prime }-n\right) }%
\left[ \theta _{1}\cos \left( \phi _{T_{S}}\left( nr\delta t\right)
/2\right) \right] \right\}   \nonumber \\
&&\times \left\{ J_{2rn^{\prime }}\left[ \theta _{2}\sin \left( r\phi
_{TL_{S}}/2\right) \right] +\left( -1\right) ^{n^{\prime }r}J_{2rn^{\prime }}%
\left[ \theta _{2}\cos \left( r\phi _{TL}/2\right) \right] \right\} ,
\label{29}
\end{eqnarray}
where $\phi _{TL_{S}}$ is a Talbot-Lau phase, defined as 
\begin{equation}
\phi _{TL_{S}}=\left( n^{\prime }-n\right) \omega _{q}T.  \label{30}
\end{equation}
It is easy to show that the Fourier component $\rho _{gg}^{(SS)}\left(
nr,\delta t=0\right) $ does not vanish for $s\neq 0$. We shall return to
this point shortly.

The remaining terms in Eq. (\ref{23}) can be evaluated in the same manner,
and one finds 
\begin{mathletters}
\label{31}
\begin{eqnarray}
\rho _{gg}\left( nr,\delta t,T\right) &=&\Phi \left( \delta t\right) F\left(
T\right) ,  \label{31a} \\
\Phi \left( \delta t\right) &=&\,^{1}/_{2}\exp \left[ -(nrqu\delta t)^{2}/4%
\right] \left\{ J_{2r\left( n^{\prime }-n\right) }\left[ \theta _{1}\sin
\left( \phi _{T_{S}}\left( nr\delta t\right) /2\right) \right] \left[
1+C\left( \phi _{T_{D}}\left( nr\delta t\right) ,-r\left( n^{\prime
}-n\right) \omega _{d}/\Gamma \right) \right] \right.  \nonumber \\
&&+\left. (-1)^{r\left( n^{\prime }-n\right) }J_{2r\left( n^{\prime
}-n\right) }\left[ \theta _{1}\cos \left( \phi _{T_{S}}\left( nr\delta
t\right) /2\right) \right] \left[ 1-C\left( \phi _{T_{D}}\left( nr\delta
t\right) ,-r\left( n^{\prime }-n\right) \omega _{d}/\Gamma \right) \right]
\right\} ,  \label{31b} \\
F\left( T\right) &=&\,^{1}/_{2}\left\{ J_{2rn^{\prime }}\left[ \theta
_{2}\sin \left( r\phi _{TL_{S}}/2\right) \right] \left[ 1+C\left( r\phi
_{TL_{D}},rn\omega _{q}/\Gamma \right) \right] \right.  \nonumber \\
&&\left. +\left( -1\right) ^{rn^{\prime }}J_{2rn^{\prime }}\left[ \theta
_{2}\cos \left( r\phi _{TL_{S}}/2\right) \right] \left[ 1-C\left( r\phi
_{TL_{D}},rn\omega _{q}/\Gamma \right) \right] \right\} ,  \label{31c}
\end{eqnarray}
where 
\end{mathletters}
\begin{equation}
\phi _{TL_{D}}=\left( n^{\prime }-n\right) \omega _{d}T.  \label{311}
\end{equation}
The result (\ref{31a}) is the product of a term, $\Phi \left( \delta
t\right) ,$ giving the time dependence of the Fourier component in the
vicinity of the echo time, and a term, $F\left( T\right) ,$ giving its
dependence on the time separation of the pulses.

We consider these terms separately, starting with $\Phi \left( \delta
t\right) $. The exponential factor in Eq. (\ref{31b}) leads to a
nonvanishing contribution to the atomic density only for times $nr\delta
t\sim (qu)^{-1}\ll \omega _{d}^{-1}<\omega _{q}^{-1}$. In this limit,
and for $\omega _{d}/\Gamma \ll 1,$ Eq. (\ref{31b}) reduces to 
\begin{equation}
\Phi \left( \delta t\right) =\exp \left[ -(nrqu\delta t)^{2}/4\right]
J_{2r\left( n^{\prime }-n\right) }\left[ \theta _{1}\phi _{T_{S}}\left(
nr\delta t\right) /2\right] .  \label{312}
\end{equation}
Comparing this expression with Eq. (\ref{227}), one sees that the atomic
density near the echo times mirrors the atomic density in the time interval $%
\delta t$ following the first pulse. Thus the dependence of the Fourier
component $s=nr$ near the echo time can be understood in terms of the
dependence of the $s$th Fourier component at a time $\delta t$ following the
first pulse. In this time interval, only those Fourier components having $%
s\lesssim \theta _{1}\omega _{q}/qu$ are created with nonnegligible
amplitude.

It is important to note that the recoil dephasing responsible for the
''washing out'' of the spontaneous contribution in the Talbot effect plays
no role here, since it is negligibly small in the time interval $\delta
t\sim (qu)^{-1}$. The recoil dephasing {\em during} the spontaneous
decay of the excited state {\em does }provide a small correction to Eq. (\ref
{312}) for $\delta t=0.$ It follows from Eq. (\ref{31b}) that, to lowest
order in $\omega _{d}/\Gamma \ll 1$ 
\begin{equation}
\Phi \left( \delta t=0\right) \approx \,^{1}/_{5}\left( -1\right) ^{r\left(
n^{\prime }-n\right) }\left[ r\left( n^{\prime }-n\right) \omega _{d}/\Gamma %
\right] ^{2}J_{2r\left( n^{\prime }-n\right) }\left( \theta _{1}\right) .
\label{321}
\end{equation}

We return now to the dependence of the Fourier components on $T$, given by $%
F\left( T\right) .$ The echo configuration considered in this section is the
same as that which leads to the Talbot-Lau effect. In the (matter wave)
Talbot-Lau effect, the atomic density is a periodic function of the time
separation between the pulses. In contrast to the normal Talbot-Lau effect,
the density (\ref{31a}) is {\em not} a periodic function of $T$, owing to
the spontaneous contributions to $F\left( T\right) $. However, for pulse
separations $T\gg \omega _{d}^{-1},$ the spontaneous processes no longer
contribute to $F(T)$ and one finds the periodic dependence 
\begin{equation}
F\left( T\right) \approx \,^{1}/_{2}\left\{ J_{2rn^{\prime }}\left[
\theta _{2}\sin \left( r\phi _{TL_{S}}/2\right) \right] +\left( -1\right)
^{rn^{\prime }}J_{2rn^{\prime }}\left[ \theta _{2}\cos \left( r\phi
_{TL_{S}}/2\right) \right] \right\} ,  \label{32}
\end{equation}
which is reminiscent of Eq. (\ref{222}). For shorter pulse separations, $%
T\ll \omega _{d}^{-1}$, the function $F(T)$ builds up as 
\begin{equation}
F\left( T\right) \approx \frac{1}{10}\left( -1\right) ^{rn^{\prime
}}J_{2rn^{\prime }}\left( \theta _{2}\right) \left[ r\left( n^{\prime
}-n\right) \omega _{d}T\right] ^{2}.  \label{33}
\end{equation}
[compare with Eq. (\ref{221})]. When the angle between ${\bf k}_{1}$ and $%
{\bf k}_{2}$ is small, one finds that, for $T\sim \omega _{d}^{-1}\ll \omega
_{q}^{-1}$, 
\begin{equation}
F\left( T\right) =\,^{1}/_{2}(-1)^{n^{\prime }r}J_{2n^{\prime }r}\left[
\theta _{2}\right] \left[ 1-C\left( r\phi _{TL_{D}},0\right) \right]
\label{34a}
\end{equation}
which is to be compared with Eq. (\ref{224}). The Talbot-Lau dependence $%
F(T) $ [See Eq. (\ref{31c})] is qualitatively similar to the Talbot
dependence of $\rho _{gg}\left( s,t\right) $ [See Eq. (\ref{17})]. In the
limit that $\Theta \ll 1$, and for $T\sim \omega _{d}^{-1}$ [ Eq. (\ref{34a}%
)] or $t\sim \omega _{d}^{-1}$ [Eq. (\ref{224})], the agreement is
quantitative.

\begin{figure}[tb!]
\centering
\begin{minipage}{8.0cm}
\epsfxsize= 8 cm \epsfysize= 5.9 cm \epsfbox{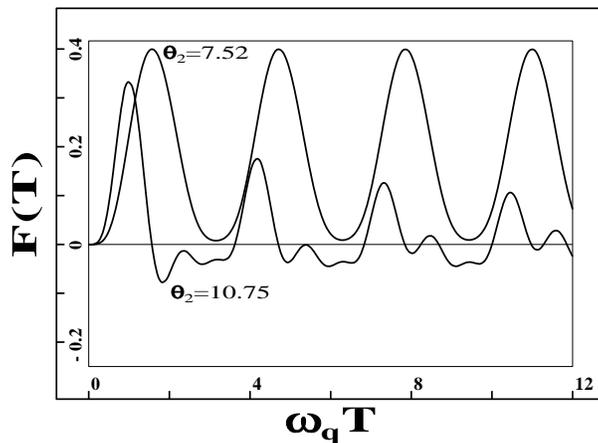}
\end{minipage}
\caption{A graph of the function $F\left( T\right) $ that gives the
dependence of the Fourier component $\protect\rho _{gg}(1,\protect\delta
t,T) $ on the separation between pulses. The pulse area $\protect\theta %
_{2}=7.52$ is chosen to maximize $F\left( T\right) $, while the area $%
\protect\theta _{2}=10.75$ is chosen to maximize the relative contribution
from spontaneous processes.}
\label{Fig3}
\end{figure}

For the case of counterpropagating waves $\left( \Theta =\pi \right) ,$ the
dependence $F\left( T\right) $ is plotted in Fig. \ref{Fig3} for $n=1$, $r=1$
($s=nr=1)$ and $n^{\prime }=2$. The pulse area $\theta _{2}=7.52$ was chosen
to maximize $F(T)$, while the area $\theta _{2}=10.75$ was chosen to
maximize the contribution from spontaneous processes. The value $\theta
_{2}=10.75$ was obtained by maximizing the ratio $F_{m}/F_{m}^{as}$, where $%
F_{m}$ is the maximum of the exact expression (\ref{31c}), which occurs at $%
\omega _{q}T\sim 1$, and $F_{m}^{as}$ is the maximum of the asymptotic
expression (\ref{32}) occurring at $\omega _{q}T\gg 1$.

\section{Summary}

We have described a new type of matter wave atom interferometer (MWAI). One
or two standing wave, resonant pulses interact with an ensemble of atoms.
Atomic motion during the pulse is neglected (Raman-Nath approximation), as
is spontaneous emission. As a result of the atom-field interactions, the
total atomic density acquires a spatial modulation that can be attributable
solely to matter-wave interference - the signals arise only for times
greater than or comparable with the inverse recoil frequency $\omega
_{d}^{-1}$ or $\omega _{q}^{-1}$. Spontaneous processes destroy the
periodicity of the Talbot or Talbot-Lau signals. However, for sufficiently
long times, the spatially modulated atomic density becomes a periodic
function of $\omega _{q}t$ (Talbot effect) or $\omega _{q}T$ (Talbot-Lau
effect).

To observe the Talbot effect discussed in Sec. III, one can use a highly
collimated ($qu\theta _{b}\ll \omega _{q})$ atomic beam that is sent through
a field interaction region. The fields can be pulsed, if necessary, to
ensure that the interaction time is much less than the excited state
lifetime. The modulated atomic density can be monitored by scattering a
probe field off the atoms or by directly depositing the atoms on a
substrate. One might also contemplate doing this experiment in the time
domain, using a Bose condensate. The Talbot-Lau effect can be observed
either in the spatial domain (using an atomic beam having an appropriate
angular divergence) or in the time domain, using a laser cooled and trapped
vapor.

Finally, we would like to comment on the fact that the Talbot-Lau Fourier
components do not vanish identically for $\delta t=0$ [see Eq. (\ref{321})].
The amplitude of these components is of order $\left( \omega _{d}/\Gamma
\right) ^{2}$, reflecting the contribution of recoil dephasing on the time
scale of the excited state lifetime. Although we did not give the equation
in the text, there is also a contribution to the Talbot Fourier components
of order $\left( \omega _{d}/\Gamma \right) ^{2}.$ These contributions
reflect the ''opening'' of the closed two-level system by the recoil
associated with spontaneous emission. As in the recoil-induced resonances 
\cite{rir}, the opening of the system is connected with quantum scattering -
it vanishes in the limit that $\hbar \sim 0$.

This situation differs from that involving phase gratings on {\em open},
two-level transitions. Imagine that the atoms have two ground states, $g$
and $g^{\prime }$, to which the excited state can decay, but that the field
drives only the $g-e$ transition. By using a far detuned field, spontaneous
emission to state $g^{\prime }$ can be suppressed by a factor ($\Gamma
/\Delta )^{2}$. Following decay, but for times much less than the inverse
recoil time, the population density $\rho _{gg}$ is spatially modulated to
order ($\Gamma /\Delta )^{2}$, as is $\rho _{g^{\prime }g^{\prime }},$ but
the total density $\left( \rho _{gg}+\rho _{g^{\prime }g^{\prime }}\right) $
is uniform. The opening of $e-g$, two-level system in this case has nothing
to do with quantum scattering effects.

\acknowledgments

We are pleased to acknowledge helpful discussions with J. L. Cohen and M.
Weitz. This research is supported by the U. S. Army Research office under
grant number DAAG5-97-0113 and by the National Science Foundation under
grants PHY-9414020 and PHY-9800981.

\end{document}